\documentclass[9pt,twocolumn,twoside]{opticajnl}
\journal{opticajournal} 

\setboolean{shortarticle}{false}

\usepackage{lineno}

\title{Physics-aware hyper-reduction for full-vector photonic waveguide mode solvers}

\author[1]{Daniel~Rodr\'iguez-Guill\'en}
\author[1,*]{Lorena~Vel\'azquez-Ibarra}

\affil[1]{Departamento de F\'isica, Divisi\'on de Ciencias e Ingenier\'ias, Universidad de Guanajuato, Le\'on 37150, M\'exico}

\affil[*]{lorenav@fisica.ugto.mx}

\begin{abstract}
Photonic waveguide design often requires repeated full-vector Maxwell eigenmode solves over wavelength and geometry, where the subpixel smoothing needed at dielectric interfaces makes operator assembly a substantial part of the cost. We present a physics-aware hyper-reduction framework for accelerating this assembly. The method exploits an exact affine dependence of the discrete operators on the locally sampled inverse-permittivity tensor, and chooses its strategy from how a parameter acts on the interface cells. We demonstrate two regimes. In wavelength sweeps at fixed geometry, tensor values change on a fixed set of cells, and proper orthogonal decomposition with matrix discrete empirical interpolation reproduces the operators from few snapshots. In moving-boundary sweeps, interface information relocates across the grid, and event-driven dynamic local sampling updates only the coefficients a geometric step changes, reproducing full assembly exactly. These results show that operator assembly can be reduced by more than an order of magnitude without approximating the moving boundary, shifting the cost of a photonic mode sweep to the eigensolver.
\end{abstract}

\setboolean{displaycopyright}{false} 

\begin{document}

\maketitle

\section{Introduction}
Photonic waveguide design relies on accurate full-vector mode solvers. Dispersion analysis, fabrication-tolerance studies, and geometry optimization require the same eigenproblem to be evaluated at many parameter values~\cite{Poletti05, Wang11, RGInverse} . The finite-difference frequency-domain (FDFD) method is attractive because it provides a direct discretization of Maxwell's equations on a structured grid~\cite{yee1966,alexopoulos2022}. However, every point in a parameter sweep requires the discrete operators to be assembled again.

Accurate operator assembly is particularly demanding near dielectric interfaces. Simple scalar or convolution-based averaging does not, in general, reproduce the correct electromagnetic interface conditions when a boundary cuts a grid cell~\cite{glytsis2018,farjadpour2006,kottke2008}. Subpixel smoothing must preserve the continuity of the normal displacement field and the tangential electric field. This treatment introduces a locally anisotropic material tensor whose values depend on the material indices, fill fractions, and interface normals~\cite{oskooi2009}. Recomputing these quantities over the full grid can make assembly a substantial part of the total simulation cost.

Projection-based hyper-reduction offers a natural strategy for accelerating this stage. Proper orthogonal decomposition (POD)~\cite{sirovich1987} and the matrix discrete empirical interpolation method (MDEIM)~\cite{chaturantabut2010,negri2015} approximate a parameter-dependent operator from a limited number of full assemblies. These methods are effective when the operator snapshots lie close to a low-dimensional linear space. Photonic parameter sweeps, however, contain two physically different types of dependence. A wavelength sweep changes dispersive material values while the geometry and interface locations remain fixed. A geometry sweep moves dielectric boundaries across the computational grid and changes which cells contain interface information.

This distinction is important because smooth coefficient variation and moving-interface variation need not have the same compressibility. In the spectral direction, the same tensor entries vary smoothly with wavelength. In the geometric direction, tensor entries are created, modified, and removed as the boundary crosses fixed Yee cells. A global linear basis can then reproduce its training geometries accurately but fail between them. Treating both parameter directions with the same reduced representation may therefore exchange assembly cost for an uncontrolled operator error. 

In this work, we show for the first time, to the best of our knowledge, that the full-vector FDFD operators depend affinely on the local inverse-permittivity tensor. We then use this structure to select the assembly strategy from the physical form of the parameter dependence. Wavelength-dependent operators are compressed with POD/MDEIM. Moving-boundary sweeps are evaluated with event-driven dynamic local sampling (edDLS), introduced here, which updates only the tensor entries that change between consecutive geometries. The incremental update is algebraically equivalent to full reassembly and does not introduce a reduced-operator approximation.

The method is evaluated using a trapezoidal ridge, a directional coupler, and a photonic crystal fiber. These examples increase the number and complexity of the moving dielectric boundaries. The numerical study compares spectral and geometric compressibility, verifies the exactness of the event update, and measures how the required local work changes across the three devices. We call the resulting framework physics-aware in a specific sense. The choice between compression and exact local updating is not made by trying both and keeping the better result, nor by inspecting snapshot spectra after the fact. It is made in advance, from a property of the parameter itself: whether varying it changes the values carried by a fixed set of interface cells, or changes which cells carry interface information at all. The first case leaves the coefficient support intact and is compressible; the second relocates it and is not. That criterion is available before any snapshot is computed, and Sections~\ref{sec:compress} and~\ref{sec:eddls} show that it predicts the observed behavior in both directions.

\section{Full-vector FDFD formulation and affine material dependence}
\label{sec:formulation}

\subsection{Full-vector generalized eigenproblem}
\label{subsec:eigenproblem}

Starting from Maxwell's equations in the frequency domain, guided-mode problems in photonic waveguides reduce to a generalized eigenvalue problem~\cite{johnson2001,alexopoulos2022}. For a source-free, time-harmonic field with $e^{-i\omega t}$ convention,
\begin{align}
    \nabla \times \mathbf{E} &= i\omega\mu_0\mathbf{H}, \label{eq:maxwell1}\\
    \nabla \times \mathbf{H} &= -i\omega\epsilon_0\epsilon_r(\lambda)\mathbf{E}, \label{eq:maxwell2}
\end{align}
where $\epsilon_r(\lambda)$ is the wavelength-dependent relative permittivity (e.g.\ from a Sellmeier model), and the media are non-magnetic. We write $\epsilon$ for $\epsilon_r$ below. For a waveguide that is uniform along the propagation direction $z$, guided modes are assumed to propagate as $\mathbf{E}(\mathbf{r}) = \mathbf{e}(x,y)e^{i\beta z}$, $\mathbf{H} (\mathbf{r}) = \mathbf{h}(x,y)e^{i\beta z}$, where $\beta$ is the propagation constant. Eliminating the longitudinal field components from Eqs.~(\ref{eq:maxwell1})--(\ref{eq:maxwell2}) yields a
curl-curl eigenmode formulation for the transverse field components.

After discretizing the transverse cross-section, the eigenmode computation becomes a large, sparse generalized eigenvalue problem,
\begin{equation}
    \mathbf{A}(\lambda, g)\,\mathbf{x}(\lambda, g) = \beta^{2}(\lambda, g)\,
    \mathbf{B}(\lambda, g)\,\mathbf{x}(\lambda, g),
    \label{eq:generalized_eig}
\end{equation}
where $\mathbf{x}$ is the discrete eigenvector containing the transverse magnetic-field degrees of freedom, $\mathbf{A}$ and $\mathbf{B}$ are sparse
matrices produced by the discretization, $\lambda$ is the wavelength (entering through $k_0(\lambda)=2\pi/\lambda$ and material dispersion), and $g$ collects the geometric design parameters that define the waveguide
cross-section. Equation~(\ref{eq:generalized_eig}) is obtained from a tensorial finite-difference frequency-domain discretization on a Yee grid~\cite{yee1966,alexopoulos2022}. The eigenvalue is $\beta^{2}$, and the  effective index reported throughout is $n_{\mathrm{eff}} = \beta/k_0$.

\subsection{Subpixel tensor construction}
\label{subsec:tensor}

Accurate treatment of a dielectric interface that does not align with the Yee grid requires more than a scalar average of $\epsilon$ inside a cut cell. Enforcing continuity of the normal displacement field $D_n$ and of the tangential electric field $E_t$ across the interface yields an effective, locally anisotropic inverse permittivity tensor~\cite{oskooi2009,kottke2008,%
farjadpour2006,alexopoulos2022},
\begin{equation}
    \overline{\overline{\tau}} \;\equiv\; \overline{\overline{\epsilon^{-1}}}
    = \left\langle \epsilon^{-1}
    \right\rangle \mathbf{P} + \left\langle \epsilon \right\rangle^{-1}
    \left(\mathbf{I} - \mathbf{P}\right),
    \label{eq:tensor_general}
\end{equation}
where $\mathbf{P} = \hat{n}\hat{n}^{T}$ projects onto the local interface normal $\hat{n}=(n_x,n_y,0)$ (no $z$ component, since interfaces are uniform along the propagation direction), and $\langle \cdot \rangle$ denotes the
simple average of the enclosed quantity over the subpixel area. We write $\overline{\overline{\tau}}$ for this smoothed inverse-permittivity tensor throughout, since it, rather than $\epsilon$ itself, is the quantity on which the discrete operators depend. Expanding \eqref{eq:tensor_general} in Cartesian components gives
\begin{align}
    \tau_{xx} &= n_x^2\left\langle \epsilon^{-1}\right\rangle
        + n_y^2 \left\langle \epsilon\right\rangle^{-1}, \label{eq:tau_xx}\\
    \tau_{yy} &= n_y^2\left\langle \epsilon^{-1}\right\rangle
        + n_x^2 \left\langle \epsilon\right\rangle^{-1}, \label{eq:tau_yy}\\
    \tau_{xy} = \tau_{yx} &= n_x n_y \left(\left\langle \epsilon^{-1}
        \right\rangle - \left\langle \epsilon\right\rangle^{-1}\right),
        \label{eq:tau_xy}\\
    \tau_{zz} &= \left\langle \epsilon \right\rangle^{-1}. \label{eq:tau_zz}
\end{align}
Defining the isotropic term $b := \langle \epsilon \rangle^{-1}$ and the anisotropy amplitude $d := \langle \epsilon^{-1} \rangle - b$, Eqs.~(\ref{eq:tau_xx})--(\ref{eq:tau_zz}) can be written compactly as
\begin{equation}
    \overline{\overline{\tau}} = b\,\mathbf{I} + d\,\hat{n}\hat{n}^{T},
    \label{eq:tau_factored}
\end{equation}
so that $\tau_{zz}=b$ and the anisotropy is carried entirely by $d$ and the orientation of $\hat{n}$. In a homogeneous cell $d=0$ and \eqref{eq:tau_factored} reduces to the scalar inverse permittivity. The continuous tensor is symmetric, so $\tau_{xy}=\tau_{yx}$. The discretization, nevertheless, stores five sampled arrays ($\tau_{xx}$, $\tau_{yy}$, $\tau_{xy}$, $\tau_{yx}$, and $\tau_{zz}$), because the two off-diagonal actions are evaluated on different Yee sublattices.

\subsection{Exact affine operator map}
\label{subsec:affine}

Once the Yee grid and the finite-difference stencils are fixed, the material-dependent entries of $\mathbf{A}$ and $\mathbf{B}$ are linear functions of the sampled tensor. This is a property of the discretization and not an approximation. The stencil weights are geometric constants of the grid, and each matrix entry is a sum of such weights multiplied by individual samples of $\tau$, so no entry contains a product, a quotient, or any other nonlinear combination of two samples. The linearity holds in $\tau$ and not in $\epsilon$: the subpixel construction of Eqs.~(\ref{eq:tau_xx})--(\ref{eq:tau_zz}) is itself nonlinear in the permittivity, which is why the smoothed inverse-permittivity tensor is the quantity carried through the rest of this work. Let $\tau_{\mathrm{bg}}(\lambda)$ denote the uniform-cladding inverse permittivity and define $\Delta\tau(\lambda,g)=\tau(\lambda,g)-\tau_{\mathrm{bg}}(\lambda)$ on the design region. Each entry $\Delta\tau_i$ represents one sampled tensor component at one Yee location and is zero wherever that component equals the uniform background. With this notation,
\begin{align}
    \mathbf{A}(\lambda, g) &= \mathbf{A}_{\mathrm{bg}}(\lambda)
        + \mathcal{L}_A\big(\Delta\tau(\lambda, g)\big), \label{eq:affine_A}\\
    \mathbf{B}(\lambda, g) &= \mathbf{B}_{\mathrm{bg}}(\lambda)
        + \mathcal{L}_B\big(\Delta\tau(\lambda, g)\big), \label{eq:affine_B}
\end{align}
where $\mathbf{A}_{\mathrm{bg}}(\lambda)$ and $\mathbf{B}_{\mathrm{bg}}(\lambda)$ contain the analytic uniform-cladding operators, including the $k_0(\lambda)^2\mathbf{I}$ term in $\mathbf{A}_{\mathrm{bg}}$. The maps $\mathcal{L}_A$ and $\mathcal{L}_B$ are fixed by the finite-difference and interpolation stencils and are independent of $\lambda$ and $g$.

Because $\mathcal{L}_A$ is linear and $\Delta\tau$ is a finite vector, the map can be written explicitly in coordinates. Let $\mathbf{e}_i$ select the $i$-th entry of $\Delta\tau$ and define
\begin{equation}
    \mathbf{S}^{A}_i := \mathcal{L}_A(\mathbf{e}_i),
    \qquad
    \mathbf{S}^{B}_i := \mathcal{L}_B(\mathbf{e}_i),
    \label{eq:affine_basis}
\end{equation}
so that Eqs.~(\ref{eq:affine_A})--(\ref{eq:affine_B}) become
\begin{equation}
    \mathbf{A}(\lambda, g) = \mathbf{A}_{\mathrm{bg}}(\lambda)
        + \sum_i \Delta\tau_i(\lambda,g)\,\mathbf{S}^{A}_i,
    \label{eq:affine_sum}
\end{equation}
and analogously for $\mathbf{B}$. Each $\mathbf{S}^{A}_i$ is the constant sparse response to a unit change of one local tensor entry. Its nonzero pattern is the stencil footprint of that entry and does not depend on $\lambda$ or $g$. The implementation applies this stencil map directly and does not form or store every $\mathbf{S}^{A}_i$ as a separate matrix.

Equation~(\ref{eq:affine_sum}) is the parameter-separable form used by projection-based hyper-reduction. Here, it is an identity of the discretization rather than an approximate representation. The practical consequence is developed in Section~\ref{sec:compress}.

At fixed wavelength, linearity gives the exact incremental identity
\begin{equation}
    \mathbf{A}(\lambda,g_{k+1}) = \mathbf{A}(\lambda,g_k)
    + \mathcal{L}_A\!\left(\tau(\lambda,g_{k+1})-\tau(\lambda,g_k)\right),
    \label{eq:affine_increment}
\end{equation}
and analogously for $\mathbf{B}$. This identity does not require $\tau(\lambda,g)$ to vary smoothly with $g$. If the difference is nonzero on only a few entries, \eqref{eq:affine_sum} shows that only their stencil contributions must be updated.
Equation~(\ref{eq:affine_increment}) is the structural fact that this paper exploits: Section~\ref{sec:compress} shows that $\tau(\lambda,g)$ is smooth in $\lambda$ but not in $g$, motivating two different treatments of
Eqs.~(\ref{eq:affine_A})--(\ref{eq:affine_B}) for the two parameter directions, developed in Sections~\ref{sec:compress} and~\ref{sec:eddls}.

\section{Spectral and geometric compression }
\label{sec:compress}

Equation~(\ref{eq:affine_sum}) fixes the sparse matrices $\mathbf{S}^{A}_i$ and places the material and geometric dependence in the coefficients $\tau_i(\lambda,g)$. The background operator retains the explicit spectral factor $k_0(\lambda)^2$. The relevant question is therefore whether the local tensor and the resulting operators can be represented by low-dimensional linear spaces. We examine this question along wavelength and geometry using the same device, discretization, and error metric.

\begin{figure}[htbp]
\centering
\includegraphics[width=\columnwidth]{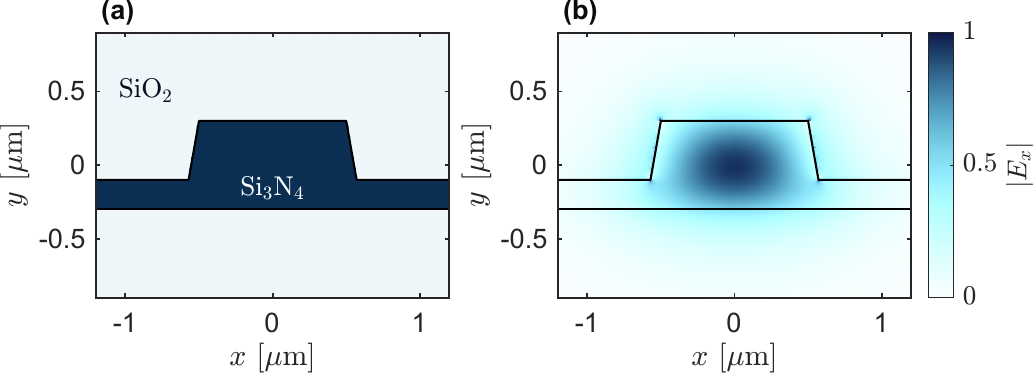}
\caption{Trapezoidal ridge used as the diagnostic problem. (a)~Refractive-index cross-section. (b)~Normalized magnitude of the dominant transverse field component fundamental quasi-TE mode at $\lambda=1.55\ \mu$m and top width $1.00\ \mu$m.}
\label{fig:ridge}
\end{figure}

The diagnostic device is the trapezoidal Si$_3$N$_4$ ridge shown in Fig.~\ref{fig:ridge}. The ridge has an $80^\circ$ sidewall on a SiO$_2$ substrate. The sloped wall produces nonzero off-diagonal tensor components and therefore tests the anisotropic interface treatment of Eqs.~(\ref{eq:tau_xx})--(\ref{eq:tau_zz}). The wavelength is swept from $1.30$ to $1.80\ \mu$m at fixed geometry. The top width is swept from $0.80$ to $1.40\ \mu$m at fixed wavelength.

The computational window is $L_x\times L_y=4.00\times2.50\ \mu\mathrm{m}^2$ and is discretized with $N_x=320$ and $N_y=270$ Yee cells, giving $dx=L_x/N_x=0.0125\ \mu$m and $dy=L_y/N_y=0.00926\ \mu$m. The generalized eigenproblem is written for the two staggered transverse magnetic-field vectors, $H_x$ and $H_y$. With the Dirichlet boundaries used here, $H_x$ contains $N_y(N_x+1)=86\,670$ samples and $H_y$ contains $(N_y+1)N_x=86\,720$ samples, for a total of $173\,390$ degrees of freedom.

\subsection{G-MDEIM baseline and held-out evaluation}
\label{subsec:baselines}

Global MDEIM (G-MDEIM) is applied directly to the assembled sparse operators. The nonzero pattern is accumulated as the union over all training points, which is necessary on a geometric axis where entries can appear or disappear as the boundary moves. The nonzero values form the snapshot matrix. POD supplies its linear basis, QDEIM selects the operator entries to evaluate, and the remaining entries are reconstructed from those samples. This is the public hyper-reduction path used in all tests below.
 
The two parameter directions use independent models. On the spectral axis, six operator snapshots from $1.30$ to $1.80\ \mu$m at intervals of $0.10\ \mu$m form the training set; the remaining $45$ points of the $0.01\ \mu$m grid form the audit set. On the geometric axis, $13$ widths separated by $0.05\ \mu$m form the training set and the remaining $48$ points of the $0.01\ \mu$m grid form the audit set. A complete FOM curve is therefore available on both axes. Modal vectors are not used to fit either model.

Accuracy is reported before the eigensolve using the relative Frobenius operator error
\begin{equation}
    \varepsilon_A(\lambda,g)=
    \frac{\|\widehat{\mathbf{A}}(\lambda,g)-\mathbf{A}(\lambda,g)\|_F}
         {\|\mathbf{A}(\lambda,g)\|_F},
    \label{eq:operator_error}
\end{equation}
and analogously for $\mathbf{B}$. Here, $\widehat{\mathbf{A}}$ denotes the operator produced by G-MDEIM and $\mathbf{A}$ is the FOM operator at the same parameter point. This comparison isolates assembly accuracy from the eigensolver.

The modal comparison uses the absolute effective-index difference
\begin{equation}
    \Delta n_{\mathrm{eff}}=
    \left|\widehat n_{\mathrm{eff}}-n_{\mathrm{eff}}^{\mathrm{FOM}}\right|,
    \label{eq:neff_error}
\end{equation}
and the residual obtained by evaluating the HR eigenpair with the FOM operators,
\begin{equation}
    r_{\mathrm{FOM}}=
    \frac{\left\|\mathbf{A}\widehat{\mathbf{x}}
    -\widehat\beta^{,2}\mathbf{B}\widehat{\mathbf{x}}\right\|_2}
    {\left\|\mathbf{A}\widehat{\mathbf{x}}\right\|_2
    +|\widehat\beta|^2\left\|\mathbf{B}\widehat{\mathbf{x}}\right\|_2}.
    \label{eq:fom_residual}
\end{equation}
Unlike the eigensolver residual evaluated on the reconstructed operators, $r_{\mathrm{FOM}}$ measures whether the HR eigenpair also satisfies the original FOM problem. Both quantities are recorded only after the selected eigenpair has been confirmed to lie on the same branch as the FOM eigenpair at the same parameter value, so that what they report is reconstruction error and never a change of modal branch.

\subsection{Wavelength dependence at fixed geometry}
\label{subsec:spectral}

Figure~\ref{fig:accuracy} reports modal accuracy, and its left column is the wavelength sweep at fixed geometry. Panel~(a) plots the effective index of the fundamental mode against wavelength for the FOM and for G-MDEIM; the two dispersion curves are visually indistinguishable, which establishes only that the reduced model stays on the physical branch and resolves nothing finer. Panel~(c) plots the effective-index discrepancy $|\Delta n_{\mathrm{eff}}|$ between them, which remains below $9.46\times10^{-7}$ across the interval and drops sharply at the training wavelengths. Panel~(e) plots the residual $r_{\mathrm{FOM}}$ of \eqref{eq:fom_residual} for the same eigenpairs, which remains below $4.93\times10^{-7}$ and reproduces the shape of panel~(c). Eigenvalue and eigenvector therefore degrade together and by comparable amounts, so the compressed operator reproduces not only the observable but the eigenproblem that defines it.

Figure~\ref{fig:operator-compressibility} reports the operator-level origin of that accuracy, and its left column is again the wavelength sweep. Panel~(a) plots the relative operator error of $\mathbf{A}$ and $\mathbf{B}$ against wavelength, which stays below $7.10\times10^{-10}$ and $1.96\times10^{-8}$ respectively. Panel~(c) plots the singular-value spectra of the two operator snapshot sets: both decay rapidly, and POD retains three directions for $\mathbf{A}$ and two for $\mathbf{B}$ out of the six training operators. The different ranks are consistent with the explicit $k_0^2=(2\pi/\lambda)^2$ contribution to $\mathbf{A}$, which is absent from $\mathbf{B}$. This is the behavior required for useful hyper-reduction: a small operator basis remains accurate between its training wavelengths.

\subsection{Geometric dependence at fixed wavelength}
\label{subsec:geometric}

The same device and implementation are then applied to the width sweep.

\begin{figure}
\centering
\includegraphics[width=\columnwidth]{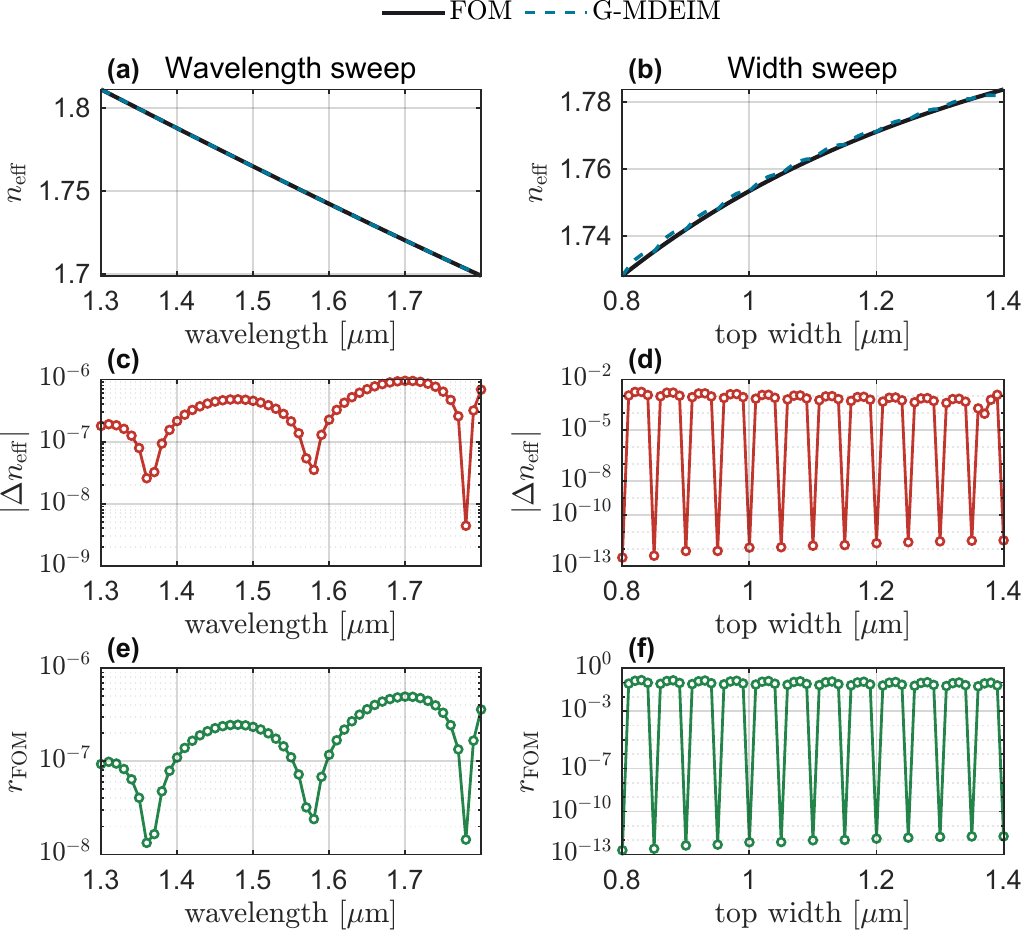}
\caption{Modal accuracy of G-MDEIM along the two parameter directions of the ridge. The left column is the wavelength sweep at fixed geometry; the right column is the top-width sweep at $\lambda=1.55\ \mu$m. (a,b)~Complete FOM and G-MDEIM curves for the fundamental-mode effective index. (c,d)~Absolute effective-index difference from \eqref{eq:neff_error}. (e,f)~Residual of the HR eigenpair evaluated with the FOM operators, \eqref{eq:fom_residual}.}
\label{fig:accuracy}
\end{figure}

The right column of Fig.~\ref{fig:accuracy} changes the parameter from wavelength to top width without changing the method. Panel~(b) shows that the G-MDEIM curve returns to the FOM curve at every training width but departs from it between successive training points. The repeated collapses in panel~(d) make this interpolation failure explicit: $|\Delta n_{\mathrm{eff}}|$ is nearly zero at the training geometries and rises again as the interface moves away from them, reaching $1.85\times10^{-3}$. Panel~(f) shows the same pattern at the eigenproblem level, where $r_{\mathrm{FOM}}$ reaches $1.48\times10^{-1}$ between training widths. A visually close effective-index curve therefore does not imply that the reconstructed eigenpair satisfies the FOM operator.

\begin{figure}
\centering
\includegraphics[width=\columnwidth]{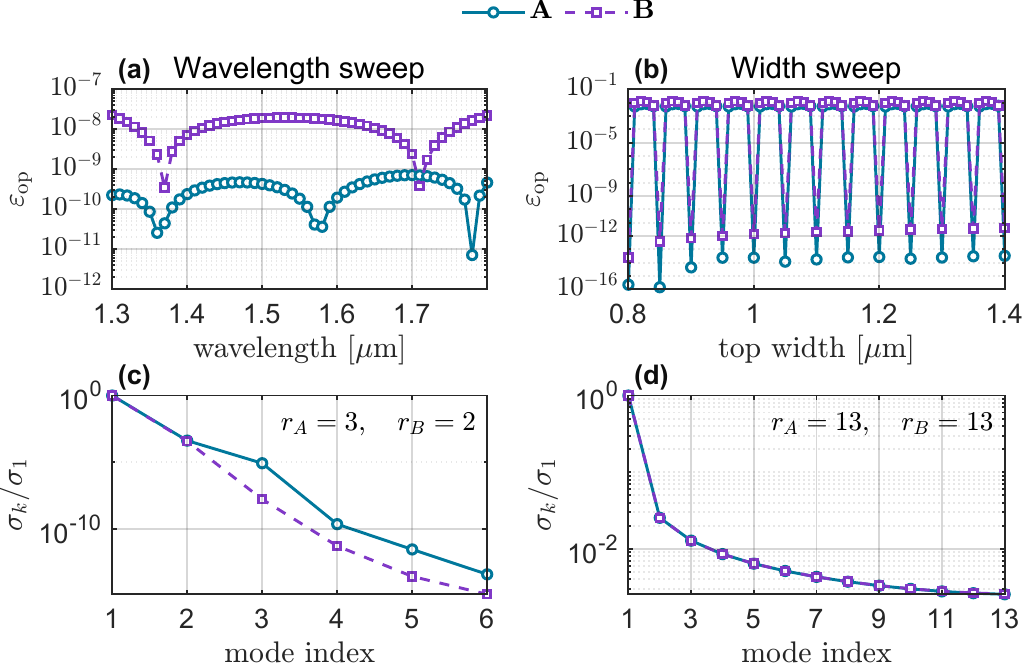}
\caption{Operator-level compression diagnostics. The left column is the wavelength sweep and the right column is the width sweep. (a,b)~Relative G-MDEIM errors of $\mathbf{A}$ and $\mathbf{B}$. (c,d)~Normalized singular spectra of the training operators. The spectral model retains only a subset of the six available snapshot directions, whereas the geometric model retains all $13$ directions and therefore does not compress its training set.}
\label{fig:operator-compressibility}
\end{figure}

Figure~\ref{fig:operator-compressibility} identifies the operator origin of this behavior. The periodic error in panel~(b) follows the $0.05\ \mu$m training interval and reaches $8.11\times10^{-3}$ for $\mathbf{A}$ and $1.28\times10^{-2}$ for $\mathbf{B}$. Unlike the rapid spectral decay in panel~(c), the geometric spectrum in panel~(d) retains all $13$ available directions for both operators. The geometric model therefore reproduces its training operators without forming a smaller space, and that full training space still fails between sampled boundary positions.

The oscillations in Fig.~\ref{fig:accuracy}(d,f) follow from the G-MDEIM operator reconstruction itself, which motivates the moving-support treatment developed in the next section.

\section{Exact event-driven dynamic local assembly}
\label{sec:eddls}

Section~\ref{sec:compress} showed that a moving interface is not represented accurately by the geometric G-MDEIM model. The affine identity of Sec.~\ref{subsec:affine} also provides an exact alternative. Instead of reconstructing the moving tensor from a global basis, we evaluate only the tensor coefficients required by the current geometric transition.

\subsection{Declared local region and geometric events}
\label{subsec:dls}

Let the active material support over a declared geometric interval $\mathcal{G}$ be
\begin{equation}
    \mathcal{R}_{\ast}=
    \bigcup_{g\in\mathcal{G}}
    \operatorname{supp}\!\left[\Delta\tau(\lambda,g)\right].
    \label{eq:design_region}
\end{equation}
A local implementation region $\mathcal{R}$ is then selected such that $\mathcal{R}_{\ast}\subseteq~\mathcal{R}$. Equality is not required. A simple rectangular region is useful in practice, provided that every Yee coefficient outside it equals the analytic background for all $g\in\mathcal{G}$. The region is therefore determined from the geometry family and its parameter bounds, not from a modal field.

Dynamic local sampling (DLS) evaluates $\tau$ on $\mathcal{R}$ and assembles the full operators through Eqs.~(\ref{eq:affine_A})--(\ref{eq:affine_B}). It introduces no basis, interpolation, or tensor reconstruction. The number of evaluated coefficients is reduced from the complete tensor storage $N_{\tau}$ to $N_{\mathcal{R}}=|\mathcal{R}|$. This ratio is a structural sampling count, and not a prediction of wall time, because uniform and interface-cut cells do not have the same evaluation cost. For the ridge discretization, $N_{\tau}=433\,771$ and the declared region contains $138\,819$ coefficients, or $32.00\%$ of the complete storage.

\begin{figure}
\centering
\includegraphics[width=\columnwidth]{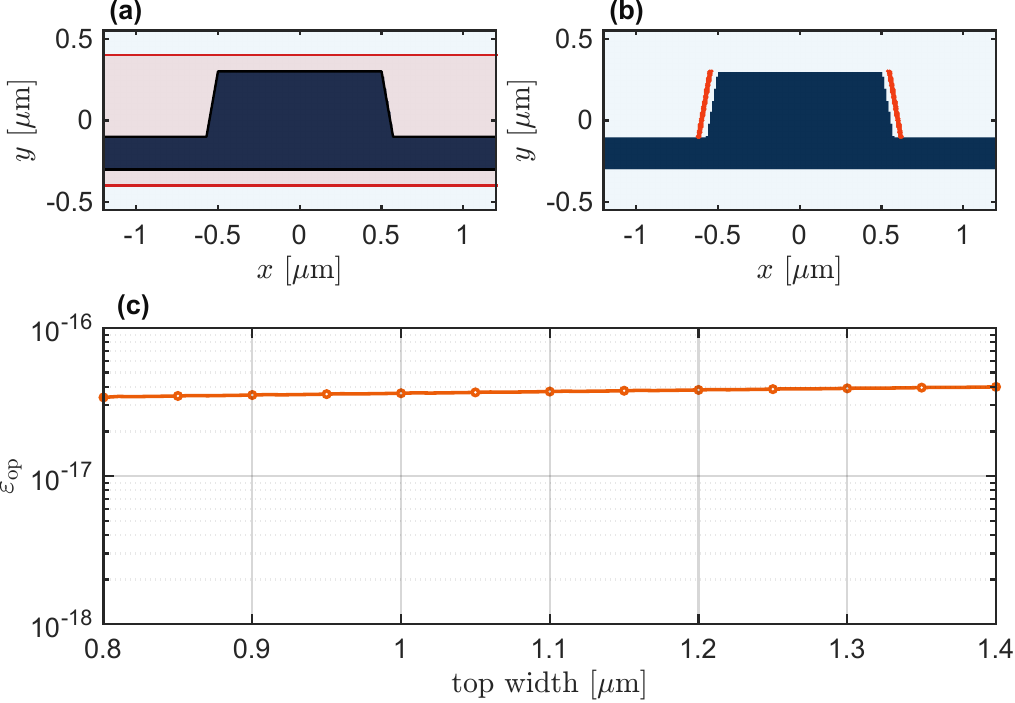}
\caption{Exact local assembly for the ridge width sweep. (a)~Conservative
DLS region $\mathcal{R}$ containing every tensor coefficient that can depart
from the analytic background. (b)~Complete spatial support for the transition
between $1.09$ and $1.10\ \mu\mathrm{m}$. Coincident tensor components are
represented by one marker per physical location. (c)~Relative operator
error $\varepsilon_{\mathrm{op}}=\max(\varepsilon_A,\varepsilon_B)$ of the
event-driven update against an independently assembled full-order operator at
all 61 widths.}
\label{fig:event}
\end{figure}

For two consecutive geometries, we define the complete event support as
\begin{equation}
    \mathcal{E}_{k}=
    \operatorname{supp}\!\left[\tau(\lambda,g_{k+1})
    -\tau(\lambda,g_k)\right].
    \label{eq:event_support}
\end{equation}
The uniform parts of the ridge and the static slab remain in $\mathcal{R}$ but do not belong to $\mathcal{E}_k$. Only the cells recut by the moving sidewalls contribute to the event. In the implementation, a coefficient is assigned to $\mathcal{E}_k$ when its change exceeds a declared numerical tolerance. The resulting index sets $\mathcal{E}_k$ are tied to the ordered parameter grid on which they were constructed. Across the 60 transitions, $|\mathcal{E}_k|$ remains between 572 and 580 coefficients. Thus, one geometric step changes $0.416\%$ of the declared region, approximately 240 times fewer coefficients than a complete DLS resampling. These counts are specific to this device and discretization; their role is to quantify the tested case, not to define the method. Figure~\ref{fig:event}(a) shows the declared region, whereas Fig.~\ref{fig:event}(b) shows the distinct Yee locations touched by one representative event. A location is displayed once, even when several tensor components change there.

\subsection{The exact incremental update}
\label{subsec:update}

At fixed wavelength, let $\delta\tau_k=\tau(\lambda,g_k)-\tau(\lambda,g_{k-1})$. By \eqref{eq:event_support}, this vector is nonzero only on $\mathcal{E}_{k-1}$. Starting from one DLS assembly at $g_1$, each subsequent operator is obtained by resampling the event coefficients and applying
\begin{equation}
    \mathbf{A}_k=\mathbf{A}_{k-1}+\mathcal{L}_A(\delta\tau_k),
    \qquad
    \mathbf{B}_k=\mathbf{B}_{k-1}+\mathcal{L}_B(\delta\tau_k).
    \label{eq:eddls_update}
\end{equation}
Only the stencil responses associated with indices in $\mathcal{E}_{k-1}$ contribute. We refer to this update as event-driven dynamic local sampling (edDLS).

The exactness follows directly from linearity. If the warm start is exact and the event support is complete, then
\begin{align}
    \mathbf{A}_k
    &=\mathbf{A}_1+\sum_{j=2}^{k}\mathcal{L}_A(\delta\tau_j) \\
    &=\mathbf{A}_1+\mathcal{L}_A\!\left(
      \tau(\lambda,g_k)-\tau(\lambda,g_1)\right)
      =\mathbf{A}(\lambda,g_k),
    \label{eq:eddls_telescoping}
\end{align}
and the same argument holds for $\mathbf{B}_k$. The result does not require a smooth dependence on $g$. In floating-point arithmetic, the remaining difference is the accumulated rounding of the sparse additions.

\begin{algorithm}[htbp]
\caption{Event-driven dynamic local sampling over an ordered grid
$g_1,\dots,g_K$ at fixed $\lambda$.}
\label{alg:eddls}
\begin{algorithmic}[1]
\Statex \textbf{Offline}, once per device and parameter grid
\State Select $\mathcal{R}$ such that $\mathcal{R}_{\ast}\subseteq\mathcal{R}$.
\State Sample $\tau$ on the design region at every $g_k$.
\State $\mathcal{E}_k \gets
       \{\,i\in\mathcal{R}:|\tau_i(g_{k+1})-\tau_i(g_k)|>\varepsilon_{\rm evt}\,\}$
       for $k=1,\dots,K-1$.
\Statex
\Statex \textbf{Online}
\State Resample $\tau(\lambda,g_1)$ on $\mathcal{R}$ and assemble
       $\mathbf{A}_1,\mathbf{B}_1$ by DLS.
\For{$k=2,\dots,K$}
  \State Resample $\tau_i(\lambda,g_k)$ for $i\in\mathcal{E}_{k-1}$ only.
  \State $\delta\tau \gets 0$;\quad
         $\delta\tau_i \gets \tau_i(\lambda,g_k)-\tau_i(\lambda,g_{k-1})$
         for $i\in\mathcal{E}_{k-1}$.
  \State Update $\mathbf{A}_k,\mathbf{B}_k$ by \eqref{eq:eddls_update}.
\EndFor
\end{algorithmic}
\end{algorithm}

Figure~\ref{fig:event}(c) audits \eqref{eq:eddls_telescoping} at every width, not only at selected points. The event-driven operator is compared with an independently assembled full-order operator at all 61 widths. The maximum operator error is $4.00\times10^{-17}$; the small variation across the curve is rounding accumulation rather than a geometric approximation.

\subsection{Scope of the exact update}
\label{subsec:eddls-scope}

edDLS exploits parametric sparsity rather than low-rank redundancy. It does not regularize, interpolate, or reconstruct a moving interface, and it does not provide an advantage when a parameter changes most tensor coefficients at every step. The current update also assumes a fixed wavelength. If $\lambda$ changes, the analytic background, the explicit $k_0^2$ term, and the dispersive material values must be updated in addition to the local geometric event.

The sets $\mathcal{E}_k$ are valid only in the declared ordered grid. A random query or a geometry outside that grid requires a new warm start and a newly certified event. Finally, the ratio $|\mathcal{E}_k|/|\mathcal{R}|$ measures the local work avoided at the coefficient level, which is not the same quantity as elapsed assembly time. The two are related by the cost model of the sampler and the update kernel, so the conversion is measured rather than inferred: Section~\ref{sec:devices} reports it directly in Table~\ref{tab:timing} for both DLS and edDLS on two devices.

\section{Hyper-reduced spectral and event-driven geometric sweeps}
\label{sec:devices}

The ridge established the two assembly mechanisms separately. We now apply the complete workflow to a directional coupler and a one-ring photonic crystal fiber. For each device, a wavelength sweep compares FOM with spectral G-MDEIM, whereas a geometric sweep compares FOM, DLS, and edDLS. Local sampling is a geometric construction: the region it restricts the sampler to is the region a moving interface can recut, so it is reported on the geometric axis only. The purpose of this section is to measure what assembly reduction is worth on a full sweep. Accuracy has already been settled on the ridge and is carried here only as an admissibility condition: a method that does not reproduce the FOM eigenpair has no cost to report. What the section adds is the separation of assembly cost from eigensolution cost, since only the first is reduced and the second bounds what any assembly method can deliver end to end.

\subsection{Devices and confirmed modal anchors}
\label{subsec:devices}

\begin{figure}[htbp]
\centering
\includegraphics[width=\columnwidth]{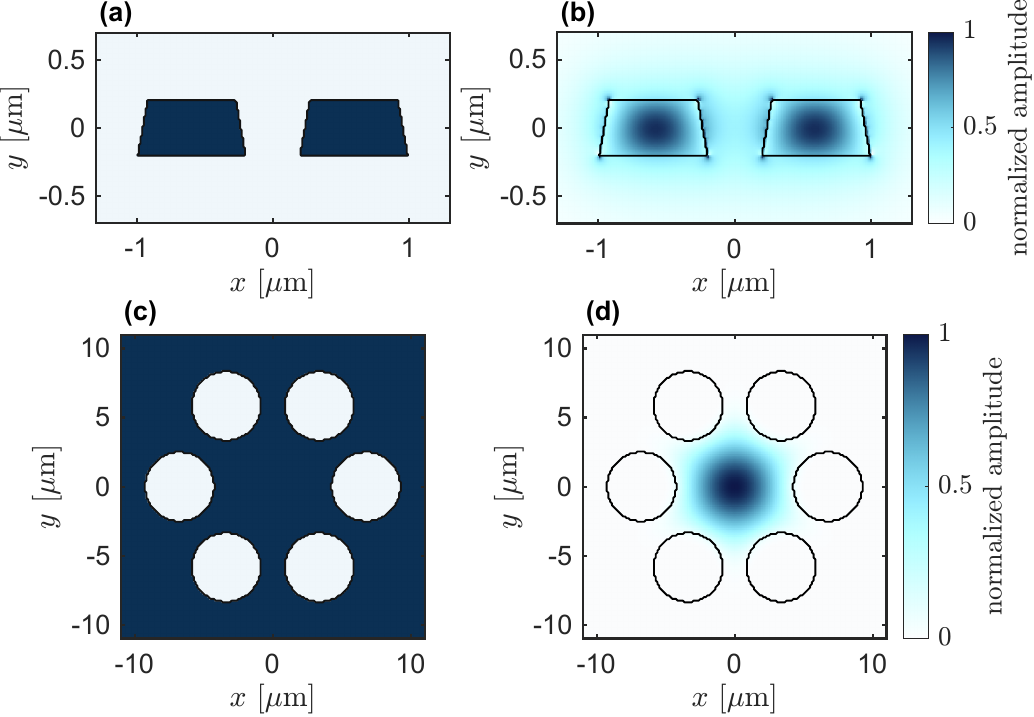}
\caption{Benchmark devices and confirmed modal anchors. (a)~Directional coupler at the reference gap. (b)~Normalized amplitude of the fundamental guided supermode used to initialize both coupler sweeps. (c)~One-ring PCF at the reference hole radius. (d)~Normalized combined amplitude of the manually confirmed fundamental polarization pair. The fields are used only to define the physical branch to be followed; the hyper-reduction models are fitted exclusively from operator snapshots.}
\label{fig:devices}
\end{figure}

Figure~\ref{fig:devices}(a) shows a directional coupler formed by two identical silicon-nitride ridges in silica. The wavelength interval is $1.20$--$1.60\ \mu\mathrm{m}$ at a reference gap of $0.40\ \mu\mathrm{m}$, and the gap interval is $0.10$--$0.70\ \mu\mathrm{m}$ at $\lambda=1.40\ \mu\mathrm{m}$. The discretization contains $300\times200$ cells and $120\,500$ magnetic-field unknowns. Before either sweep is evaluated, the fundamental guided supermode in Fig.~\ref{fig:devices}(b) is selected from the reference eigenspectrum.

The PCF in Fig.~\ref{fig:devices}(c) has six circular air holes with a pitch of $6.75\ \mu\mathrm{m}$. Its wavelength interval is $1.30$--$1.60\ \mu\mathrm{m}$ at a reference hole radius of $2.50\ \mu\mathrm{m}$, and its radius interval is $2.20$--$2.80\ \mu\mathrm{m}$ at $\lambda=1.45\ \mu\mathrm{m}$. The $240\times240$ discretization contains $115\,680$ magnetic-field unknowns. Because the fundamental core mode forms a nearly degenerate polarization pair, the two-dimensional eigenspace is selected using core confinement and then confirmed manually. Figure~\ref{fig:devices}(d) shows the rotationally-invariant amplitude formed from that pair.

Here, greater geometric complexity means that one parameter displaces more interfaces over a more distributed part of the grid. It does not indicate a different material law or more complicated electromagnetic physics. In particular, changing the coupler gap moves four sidewall branches, whereas changing the PCF radius moves six closed boundaries. These cases test whether the event construction remains practical when the moving support is no longer concentrated around one ridge.

\subsection{Fine-grid modal continuation}
\label{subsec:modal-continuation}

Mode identification is kept separate from operator reduction. Starting from the confirmed coupler field $\mathbf{x}_{k-1}$, the candidate at the next point is selected by the normalized vector overlap
\begin{equation}
 \rho_k=\max_j\frac{|\mathbf{x}_{k-1}^{H}\mathbf{x}_{k,j}|}
 {\|\mathbf{x}_{k-1}\|_2\,\|\mathbf{x}_{k,j}\|_2}.
 \label{eq:vector-overlap}
\end{equation}
For the PCF, individual eigenvectors can rotate inside the nearly degenerate fundamental space. If $\mathbf{Q}_{k-1}$ and $\mathbf{Q}_{k,j}$ are orthonormal bases for the previous and candidate two-dimensional spaces, respectively, we instead maximize
\begin{equation}
 \rho_k^{\mathrm{sub}}=\sigma_{\min}\!\left(
 \mathbf{Q}_{k-1}^{H}\mathbf{Q}_{k,j}\right).
 \label{eq:subspace-overlap}
\end{equation}
The wavelength increment is $0.005\ \mu\mathrm{m}$ for both devices. The coupler-gap increment is $0.005\ \mu\mathrm{m}$ and the PCF-radius increment is $0.010\ \mu\mathrm{m}$. These fine grids do not improve the operator approximation by themselves; they make the physical branch vary sufficiently smoothly for Eqs.~(\ref{eq:vector-overlap}) and~(\ref{eq:subspace-overlap}) to provide an auditable continuation criterion.

\subsection{Directional-coupler sweeps}
\label{subsec:coupler-sweeps}

\begin{figure}
\centering
\includegraphics[width=\columnwidth]{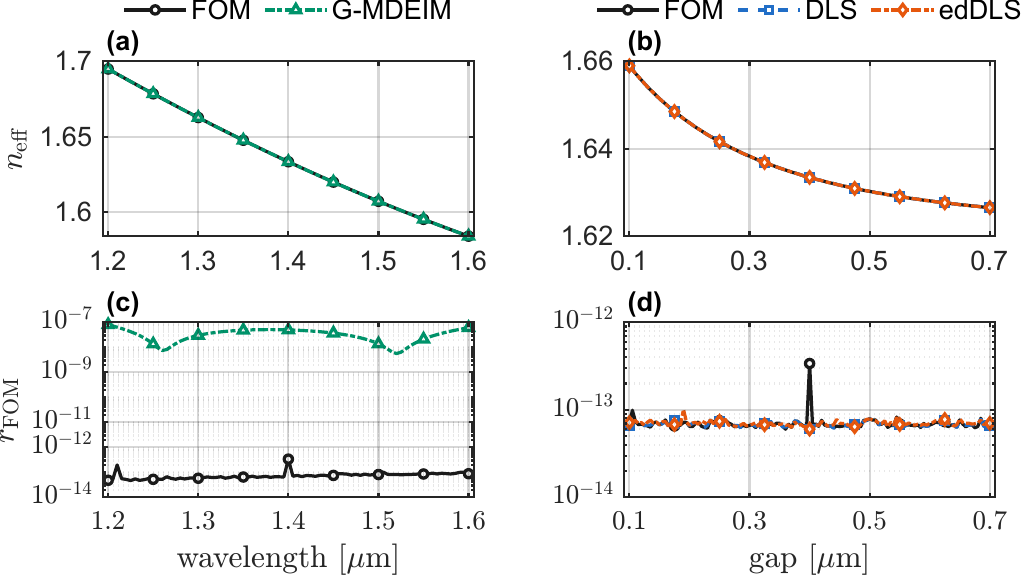}
\caption{Directional-coupler sweeps. Left column: wavelength at fixed gap. Right column: gap at fixed wavelength. (a,b) Effective index. (c,d) Residual of each eigenpair evaluated with the FOM operators. The assembly paths are FOM and G-MDEIM for wavelength, and FOM, exact DLS and edDLS for geometry.}
\label{fig:coupler-sweeps}
\end{figure}

Figure~\ref{fig:coupler-sweeps} places the wavelength and gap sweeps side by side. The first row compares the complete effective-index curves. The second row evaluates every accelerated eigenpair with the independently assembled FOM operator through \eqref{eq:fom_residual}. Along wavelength, G-MDEIM is the reduced operator model; along gap, edDLS is the exact incremental update and DLS is the exact local-assembly path it refines.

The two axes therefore carry different comparisons, because the two mechanisms answer different obstructions. On wavelength, the interface is fixed and the operator is compressible, so the question is how far a trained model can be trusted between its training points. On gap, the interface moves and the operator is not compressible, so the question is instead how little of the tensor a step actually disturbs.

\subsection{Photonic-crystal-fiber sweeps}
\label{subsec:pcf-sweeps}

\begin{figure}[htbp]
\centering
\includegraphics[width=\columnwidth]{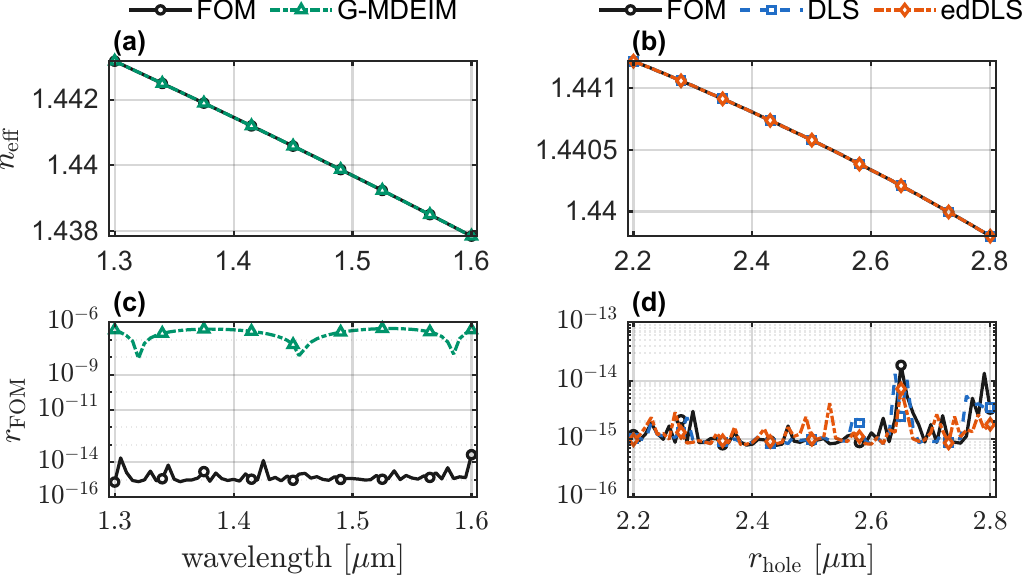}
\caption{PCF sweeps of the confirmed fundamental polarization subspace. Left column: wavelength at fixed hole radius. Right column: radius at fixed wavelength. (a,b)~Mean effective index of the pair. (c,d)~FOM residual.}
\label{fig:pcf-sweeps}
\end{figure}

Figure~\ref{fig:pcf-sweeps} repeats the same audit for the PCF. The reported effective index is the mean of the two values in the tracked fundamental polarization pair. Equations~(\ref{eq:vector-overlap}) and~(\ref{eq:subspace-overlap}) are used internally to maintain the branch, while the figure retains only the physical curve and its FOM residual. This avoids giving visual weight to an internal tracking diagnostic once the branch has been verified.

\subsection{Assembly cost}
\label{subsec:assembly-cost}

Accuracy and timing are generated in separate runs. The benchmark randomizes the method order and evaluates each full sweep once after the machine is declared idle. The reported seconds are therefore representative measurements on the present platform rather than hardware-independent performance constants.

Two times are accumulated over a sweep: the assembly time $t_{\mathrm{asm}}$ spent constructing $\mathbf{A}$ and $\mathbf{B}$, and the eigensolution time spent in the sparse eigensolver, which also carries the branch selection of Eqs.~(\ref{eq:vector-overlap}) and~(\ref{eq:subspace-overlap}). Writing $t_{\mathrm{tot}}$ for their sum, each accelerated path is compared with the full-order path over the same sweep by
\begin{equation}
    S_{\mathrm{asm}} =
    \frac{t_{\mathrm{asm}}^{\mathrm{FOM}}}{t_{\mathrm{asm}}^{\mathrm{HR}}},
    \qquad
    S_{\mathrm{tot}} =
    \frac{t_{\mathrm{tot}}^{\mathrm{FOM}}}{t_{\mathrm{tot}}^{\mathrm{HR}}}.
    \label{eq:speedups}
\end{equation}
$S_{\mathrm{asm}}$ measures the stage this work reduces; $S_{\mathrm{tot}}$ measures what a sweep user observes. The two differ because the eigensolution is untouched, and it is what bounds the end-to-end gain once assembly has been removed.

\begin{figure}
\centering
\includegraphics[width=\columnwidth]{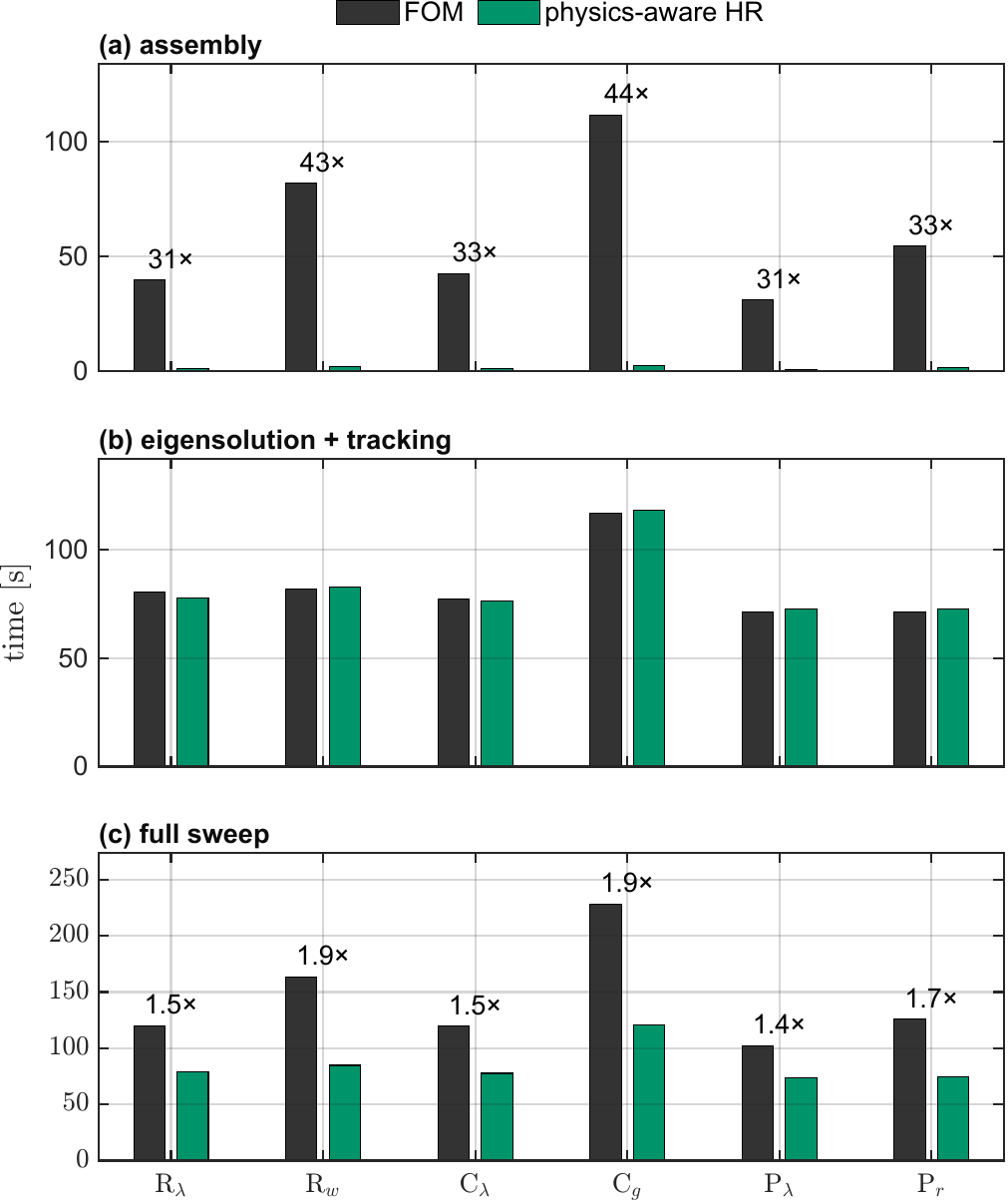}
\caption{Where the time of a full sweep goes, before and after hyper-reduction, for the six sweeps of the three devices. (a)~Assembly. (b)~Eigensolution and tracking. (c)~Full sweep, the sum of the two. Here R, C and P denote the ridge, the coupler and the PCF; the subscripts identify the wavelength, top-width, gap and hole-radius sweeps. The accelerated path is G-MDEIM for the wavelength sweeps and edDLS for the geometric ones. Each panel carries its own vertical scale. Panels~(a) and~(c) are annotated with the ratio of the two bars; panel~(b) is not, because there the two paths solve the same eigenproblem and the ratio is one by construction. Assembly collapses on every device, the eigensolution is untouched, and the full sweep therefore improves only by the share that assembly used to occupy.}
\label{fig:cost-shift}
\end{figure}

Figure~\ref{fig:cost-shift} reports the result for all three devices at once, and Table~\ref{tab:timing} gives the times behind it. Panel~(a) is the claim of this work: on every device and on both axes, the bar for hyper-reduced assembly collapses against the full-order bar. The reduction is of the same order on the diagnostic ridge as on the two larger devices, which is what a mechanism acting on the tensor rather than on the device should do. Table~\ref{tab:timing} separates, on the geometric axis, two effects that the panel shows only in combination. Confining the sampler to the region a moving interface can recut already removes most of the full-grid work, with no training and no approximation. Restricting the update further to the coefficients that a single increment actually changes then removes a large part of what remains. What the event restriction adds on top of local sampling falls monotonically as the moving geometry spreads out: it is largest on the ridge, whose two sidewalls are the most concentrated case, smaller on the four sidewall branches of the coupler, and smallest on the six closed boundaries of the PCF. That ordering is the one predicted by the event fraction of Section~\ref{sec:eddls}, and it is the expected limitation of the mechanism rather than a defect of the implementation: an increment that disturbs more of the grid leaves less for an event to skip.

\begin{table}[htbp]
\centering
\caption{Assembly time accumulated over each full sweep, $t_{\mathrm{asm}}$, and the two speed-ups of \eqref{eq:speedups}: $S_{\mathrm{asm}}$ over the assembly stage alone and $S_{\mathrm{tot}}$ over the full sweep. The first row of each sweep is the full-order path against which both ratios are formed, so it carries no ratio of its own. Values come from one controlled execution of each sweep on the present platform.}
\label{tab:timing}
\begin{tabular}{llrrr}
\hline\hline
Sweep & Method & $t_{\mathrm{asm}}$ [s] & $S_{\mathrm{asm}}$ & $S_{\mathrm{tot}}$ \\
\hline
\multicolumn{5}{l}{Trapezoidal ridge}\\
Wavelength & FOM     & $39.65$  & ---    & ---   \\
           & G-MDEIM & $1.30$   & $30.5$ & $1.52$ \\
Top width  & FOM     & $81.81$  & ---    & ---   \\
           & DLS     & $3.63$   & $22.5$ & $1.92$ \\
           & edDLS   & $1.90$   & $43.0$ & $1.93$ \\
\hline
\multicolumn{5}{l}{Directional coupler}\\
Wavelength & FOM     & $42.58$  & ---    & ---   \\
           & G-MDEIM & $1.30$   & $32.7$ & $1.54$ \\
Gap        & FOM     & $111.57$ & ---    & ---   \\
           & DLS     & $3.99$   & $28.0$ & $1.88$ \\
           & edDLS   & $2.54$   & $43.9$ & $1.89$ \\
\hline
\multicolumn{5}{l}{Photonic crystal fiber}\\
Wavelength & FOM     & $31.08$  & ---    & ---   \\
           & G-MDEIM & $0.99$   & $31.4$ & $1.39$ \\
Radius     & FOM     & $54.57$  & ---    & ---   \\
           & DLS     & $2.50$   & $21.8$ & $1.69$ \\
           & edDLS   & $1.65$   & $33.1$ & $1.69$ \\
\hline\hline
\end{tabular}
\end{table}

Panels~(b) and~(c) place that reduction in context rather than letting it stand alone. Eigensolution and tracking are unchanged, because both assembly paths hand the same-size generalized eigenproblem to the same sparse eigensolver; for the exact methods the two operators are numerically identical, so any residual difference in panel~(b) is measurement noise and not a property of the method. Panel~(c) is the consequence: once assembly has collapsed, it occupies only a small share of the online time, and the full-sweep gain is bounded by the share it used to occupy. The bottleneck has moved from operator construction to eigensolution.

Once assembly is sufficiently inexpensive, the full-order eigensolve becomes the natural next bottleneck. Projection-based eigenvector continuation can reduce that complementary stage~\cite{Rodriguez26}, so the two reductions can in principle be combined. That coupled formulation is outside the scope of the present work, which isolates the decomposition and assembly of the full-vector operators.

\section{Conclusion}

We introduced a physics-aware hyper-reduction strategy for full-vector FDFD mode solvers. The starting point is an exact affine dependence of the discrete operators on the locally sampled inverse-permittivity tensor. This structure separates two parameter variations that should not be treated identically. At fixed geometry, wavelength dependence is smooth and is represented efficiently by POD/MDEIM. When a dielectric boundary moves, the tensor support relocates across the Yee grid and a global linear basis can lose accuracy between its training geometries.

For moving boundaries, edDLS replaces global reconstruction with exact incremental updates supported only on the coefficients changed by the current geometric event. The resulting operators agree with fresh DLS and FOM assembly to numerical precision. Tests on a trapezoidal ridge, a directional coupler, and a one-ring PCF show that the same construction remains effective as the number and spatial distribution of moving interfaces increase. In the controlled sweeps, the measured assembly reductions are approximately $30$-$44\times$ while the accelerated eigenpairs retain FOM-level residuals.

These results also expose the next computational limit. Once assembly has been reduced to a small fraction of the sweep time, the unreduced sparse eigensolver dominates the calculation. Reducing that complementary stage can be pursued independently and combined with the present operator treatment. The contribution of this work is the assembly component: a method that chooses compression or exact local updating according to how the physical parameter modifies the discrete material tensor.

\begin{backmatter}
\bmsection{Funding} 
Secretaría de Ciencia, Humanidades, Tecnología e Innovación (SECIHTI, Ref.~CBF-2025-I-2037). 

\bmsection{Acknowledgment} 
D.R.-G. acknowledges financial support from SECIHTI (Graduate Fellowship, CVU-1142156).

\bmsection{Disclosures} 
The authors declare no conflicts of interest.

\bmsection{Data availability} Data underlying the results presented in this paper are not publicly available at this time but may be obtained from the authors upon reasonable request.

\end{backmatter}

\bigskip

\bibliography{refs}
\bibliographyfullrefs{refs}

\end{document}